\documentclass[twocolumn]{aastex63}
\pdfoutput=1 
\usepackage{amsmath,amstext}
\usepackage{graphicx,multirow,graphbox}

\def\ra#1#2#3{#1$^{\rm h}$#2$^{\rm m}$#3$^{\rm s}$}
\def\dec#1#2#3{#1$^\circ$#2$'$#3$''$}

\shorttitle{A Late-Time Galaxy-Targeted Search for the Radio Counterpart of GW190814}
\shortauthors{Alexander et al.}

\begin{document}

\title{A Late-Time Galaxy-Targeted Search for the Radio Counterpart of GW190814}

\author[0000-0002-8297-2473]{K.~D.~Alexander}
\altaffiliation{NHFP Einstein Fellow}\affiliation{Center for Interdisciplinary Exploration and Research in Astrophysics (CIERA) and Department of Physics and Astronomy, Northwestern University, Evanston, IL 60208, USA}

\author[0000-0001-9915-8147]{G.~Schroeder}
\affiliation{Center for Interdisciplinary Exploration and Research in Astrophysics (CIERA) and Department of Physics and Astronomy, Northwestern University, Evanston, IL 60208, USA}

\author[0000-0001-8340-3486]{K. Paterson}
\affiliation{Center for Interdisciplinary Exploration and Research in Astrophysics (CIERA) and Department of Physics and Astronomy, Northwestern University, Evanston, IL 60208, USA}

\author[0000-0002-7374-935X]{W. Fong}
\affiliation{Center for Interdisciplinary Exploration and Research in Astrophysics (CIERA) and Department of Physics and Astronomy, Northwestern University, Evanston, IL 60208, USA}

\author[0000-0002-2478-6939]{P.~Cowperthwaite}
\altaffiliation{NHFP Hubble Fellow}\affiliation{The Observatories of the Carnegie Institution for Science, 813 Santa Barbara Street, Pasadena, CA 91101, USA}

\author[0000-0001-6395-6702]{S.~Gomez}
\affiliation{Center for Astrophysics {\textbar} Harvard \& Smithsonian, 60 Garden St., Cambridge, MA 02138, USA}

\author[0000-0001-8405-2649]{B.~Margalit}
\altaffiliation{NHFP Einstein Fellow}
\affiliation{Astronomy Department and Theoretical Astrophysics Center, University of California, Berkeley, Berkeley, CA 94720, USA}

\author[0000-0003-4768-7586]{R. Margutti}
\affiliation{Center for Interdisciplinary Exploration and Research in Astrophysics (CIERA) and Department of Physics and Astronomy, Northwestern University, Evanston, IL 60208, USA}

\author[0000-0002-9392-9681]{E.~Berger}
\affiliation{Center for Astrophysics {\textbar} Harvard \& Smithsonian, 60 Garden St., Cambridge, MA 02138, USA}

\author[0000-0003-0526-2248]{P.~Blanchard}
\affiliation{Center for Interdisciplinary Exploration and Research in Astrophysics (CIERA) and Department of Physics and Astronomy, Northwestern University, Evanston, IL 60208, USA}

\author[0000-0002-7706-5668]{R. Chornock}
\affiliation{Center for Interdisciplinary Exploration and Research in Astrophysics (CIERA) and Department of Physics and Astronomy, Northwestern University, Evanston, IL 60208, USA}

\author[0000-0003-0307-9984]{T.~Eftekhari}
\affiliation{Center for Astrophysics {\textbar} Harvard \& Smithsonian, 60 Garden St., Cambridge, MA 02138, USA}

\author[/0000-0003-1792-2338]{T.~Laskar}
\affiliation{Department of Physics, University of Bath, Claverton Down, Bath BA2 7AY, UK}

\author[0000-0002-4670-7509]{B.~D.~Metzger}
\affiliation{Department of Physics and Columbia Astrophysics Laboratory, Columbia University, New York, NY 10025, USA}
\affiliation{Center for Computational Astrophysics, Flatiron Institute, 162 5th Ave, NY 10011, USA}

\author[0000-0002-2555-3192]{M.~Nicholl}
\affiliation{Birmingham Institute for Gravitational Wave Astronomy and School of Physics and Astronomy, University of Birmingham, Birmingham B15 2TT, UK}

\author[0000-0002-5814-4061]{V.~A.~Villar}
\altaffiliation{Simons Junior Fellow}
\affiliation{Department of Astronomy, Columbia University, New York, NY 10027-6601, USA}

\author[0000-0003-3734-3587]{P. K. G. Williams}
\affiliation{Center for Astrophysics {\textbar} Harvard \& Smithsonian, 60 Garden St., Cambridge, MA 02138, USA}
\affiliation{American Astronomical Society, 1667 K St. NW Ste. 800, Washington, DC 20006, USA}

\begin{abstract}
GW190814 was a compact object binary coalescence detected in gravitational waves by Advanced LIGO and Advanced Virgo that garnered exceptional community interest due to its excellent localization and the uncertain nature of the binary's lighter-mass component (either the heaviest known neutron star, or the lightest known black hole). Despite extensive follow up observations, no electromagnetic counterpart has been identified. Here we present new radio observations of 75 galaxies within the localization volume at $\Delta t\approx 35-266$ days post-merger. Our observations cover $\sim32$\% of the total stellar luminosity in the final localization volume and extend to later timescales than previously-reported searches, allowing us to place the deepest constraints to date on the existence of a radio afterglow from a highly off-axis relativistic jet launched during the merger (assuming that the merger occurred within the observed area). For a viewing angle of $\sim46^{\circ}$ (the best-fit binary inclination derived from the gravitational wave signal) and assumed electron and magnetic field energy fractions of $\epsilon_e=0.1$ and $\epsilon_B=0.01$, we can rule out a typical short gamma-ray burst-like Gaussian jet with {an opening angle of $15^{\circ}$ and} isotropic-equivalent kinetic energy $2\times10^{51}$ erg propagating into a constant density medium ${n\gtrsim0.1}$ cm$^{-3}$. These are the first limits resulting from a galaxy-targeted search for a radio counterpart to a gravitational wave event, and we discuss the challenges, and possible advantages, of applying similar search strategies to future events using current and upcoming radio facilities.
\end{abstract}


\section{Introduction}

Recent detections of gravitational waves (GW) have revolutionized our understanding of the population of compact object binaries, impacting many areas of physics and astrophysics \citep{gw150914,lvc170817,O1O2Catalog,O1O2CatalogBBH,lvc190814,O3aCatalog,populations20}. While much can be learned from the GW signals alone, understanding the full astrophysical context of the merger event, including the association to its host galaxy, requires discovery of an electromagnetic (EM) counterpart. Mergers of two neutron stars are thus of particular interest, as they are predicted to produce radiation across the EM spectrum and have been long-theorized to be the origin of short gamma-ray bursts (SGRBs; e.g.~\citealt{Narayan92, Eichler89, fong13, ber14}). This was spectacularly confirmed by the discovery of the binary neutron star (BNS) merger GW170817 \citep{lvc170817,em170817}, which not only had associated gamma-ray emission \citep{fermi170817,GW170817grb,sfk+17}, but also a bright kilonova detected in the ultraviolet, optical, and IR bands \citep{Andreoni17,Arcavi17,chornock17,Cowperthwaite17,Coulter17,Diaz17,Drout17,Kasliwal17,Nicholl17,Lipunov17,Pian17,Pozanenko17,Smartt17,Tanvir17,Utsumi17,Valenti17,Villar17,Villar18} and a long-lasting synchrotron afterglow detected from radio through X-ray wavelengths \citep{alex17,alex18, Haggard17, hal17, marg17,marg18, Troja17,Troja18,Troja19,Troja20, dav18,dob18,lyman18,mool18,Mooley18b,Mooley18c,Nynka18,Ruan18,fong19,Ghirlanda+19,Hajela19,lamb19,piro19}.

Mergers between a neutron star and a black hole are also predicted to result in detectable EM emission in some cases, particularly if the mass ratio of the binary is not too extreme \citep{Kawaguchi16,Metzger19}. It is however an open question whether neutron star-black hole (NSBH) mergers also produce SGRBs \citep{MurguiaBerthier2017b,Gompertz20}, and how those SGRBs would compare to the cosmological SGRB population. The prompt gamma-ray emission from off-axis relativistic jets is highly suppressed due to relativistic beaming, and would likely be undetectable at the larger distances where most GW mergers will occur \citep{GW170817grb,fermi170817,mc20rev}. Thus, except for the small fraction of on-axis mergers, the best opportunity to determine whether NSBH mergers produce relativistic jets or outflows of sub-relativistic material is to search for synchrotron emission in the radio or X-ray, as was done for GW170817.

On 2019 August 14, Advanced LIGO/Virgo reported the detection of a new compact object merger candidate GW190814, with a preliminary false alarm rate of one in $10^{25}$ years (GCN 25324; \citealt{gcn25324}). It was initially classified as a MassGap event (meaning that the lighter member of the binary had a mass between $3-5M_{\odot}$), but the classification was revised to a NSBH merger approximately 12 hours later (GCN 25333; \citealt{gcn25333}). This classification, together with the excellent localization (23 deg$^2$ with 90\% confidence in the skymap provided by \citealt{gcn25333} 13.5 hours post-merger), generated considerable interest and telescope investment from the astronomy community. Numerous follow up efforts across the EM spectrum revealed no evidence for any counterpart \citep{Dobie2019,Gomez2019,Ackley2020,Andreoni2020,Antier20,Gompertz20,Morgan2020,Page2020,Thakur20,Viera20,Watson20}, broadly consistent with the highly unequal binary mass ratio revealed by the full gravitational wave analysis \citep{lvc190814} and the small NS radius inferred from observations of GW170817 \citep{Capano+20}. The nature of the lighter $2.59^{+0.08}_{-0.09}M_{\odot}$ component -- neutron star or black hole -- remains unclear \citep{populations20,el20,tso+20,Tews+21}. Nevertheless, GW190814 provided an excellent test-bed for various multi-wavelength observing strategies, as its precise localization (tightened to 18.5 deg$^2$ in the final analysis by \citealt{lvc190814}) and large distance ($241^{+41}_{-45}$ Mpc) will likely be typical of GW events discovered in O4 and beyond. 

In particular, GW190814 prompted several independent searches for a radio counterpart. Unlike the optical sky, the variable radio sky is not well-characterized on timescales of months at the typical flux densities of plausible gravitational-wave transients (although the background rate of extragalactic radio transients is expected to be low, e.g. \citealt{metzger15}). {This is largely a technological limitation: high-resolution radio imaging generally requires an interferometer and most radio interferometers have very small fields of view. Thus, early wide-field radio transient searches were often shallow (with sensitivity limits $>$ few mJy) and had very limited temporal coverage, relying on just a few epochs to assess variability (e.g.~\citealt{gt86,lev02,thy11,Hodge+13}). Alternate strategies were to obtain repeated deep observations of a small region of sky (e.g. \citealt{frail94,Carilli+03,Ofek11,Mooley13,Hancock+16}), or to repurpose data originally taken for other purposes (e.g. calibration fields, \citealt{Bower07,Bell11,frail12}). In all cases, the number of highly variable sources discovered was small, limiting the conclusions that could be drawn.} 

{More recently, improvements in the mapping speed of existing radio facilities like NSF's Karl G. Jansky Very Large Array (the VLA) and the advent of new facilities with larger fields of view like ASKAP have enabled a new generation of sensitive wide-field radio surveys, refining estimates of the occurrence rates of radio transients on timescales of days to years \citep{Mooley16,mooley19,bbm+18}. Comparison of new data to previous all-sky surveys has also begun to probe the population of radio transients and variables on timescales of decades \citep{Law18, Mooley16, nyland20, wolo21}. These searches confirm earlier results that only a few percent of unresolved radio sources are highly variable on these timescales; most of the variables are active galactic nuclei (AGN; \citealt{Mooley16,bbm+18,Radcliffe19}). The typical amplitude of AGN variability is small (few percent to factors of $\sim$few, e.g.~\citealt{hov08,Richards+11,sarb20}), but extreme variability on decadal timescales is possible; for example, \cite{nyland20} recently discovered a population of quasars that have increased in brightness by 100\% to $>2500$\% on timescales of $\lesssim20$ yr.} 

While further work remains to fully characterize the variable radio sky on the timescales most relevant for GWs (months to years), the low background of transient and variable sources revealed by these searches suggests that GW counterpart searches in the radio sky may be promising. Several wide-field radio searches for GW merger counterparts have been previously employed even in cases when no bright radio counterpart is expected, to better understand the likely background rates of potential contaminating sources (e.g.~\citealt{mooley18d, bhakta20}). However, GW190814 was the first event for which a significant fraction of the localization area could be covered to any significant depth by current radio facilities. A wide-field single-frequency radio search covering 89\% of the \cite{gcn25333} localization region was conducted with ASKAP at early times ($2-33$ days post-merger), ruling out the presence of an on-axis relativistic jet with isotropic-equivalent kinetic energy $E_{\rm iso}=10^{51}$ erg within the observed region under standard assumptions about the jet microphysics \citep{Dobie2019}. This is consistent with the lack of bright X-ray or gamma-ray emission observed at early times, which would have been expected if such a jet were present \citep{gcn25341,Page2020,Watson20}.

Here, we present targeted late-time radio observations of 75 galaxies within GW190814's localization volume, spanning $1-7$ months post-discovery. All observations were taken with the NSF's Karl G.~Jansky Very Large Array (the VLA). Our data are aimed at constraining the presence of highly off-axis initially-relativistic jets, which might be expected in GW190814 given the measured high inclination of the system from the GW signal ($46^{+17}_{-11} \deg$; \citealt{lvc190814}). The timescale of our observations also allows for some limited constraints on the presence of slower-moving kilonova (KN) ejecta. A secondary goal is to characterize the background of variable and transient radio sources likely to be encountered in future radio searches for gravitational wave counterparts, with a focus on the implications for galaxy-targeted strategies. We present our observations in Section \ref{sec:obs} and discuss our counterpart search and additional targeted observations of our most promising candidate in Section \ref{sec:search}, ultimately concluding that this object is more likely to be an unrelated background source. In Section \ref{sec:disc}, we place limits on the existence of a radio counterpart to GW190814 and discuss the nature of the unrelated variable radio sources uncovered in our search. We conclude in Section \ref{sec:conc} with some implications of our work for future radio searches for GW counterparts. 

\section{Observations}\label{sec:obs}

\begin{figure*}
$\vcenter{\hbox{\includegraphics[width=0.46\textwidth]{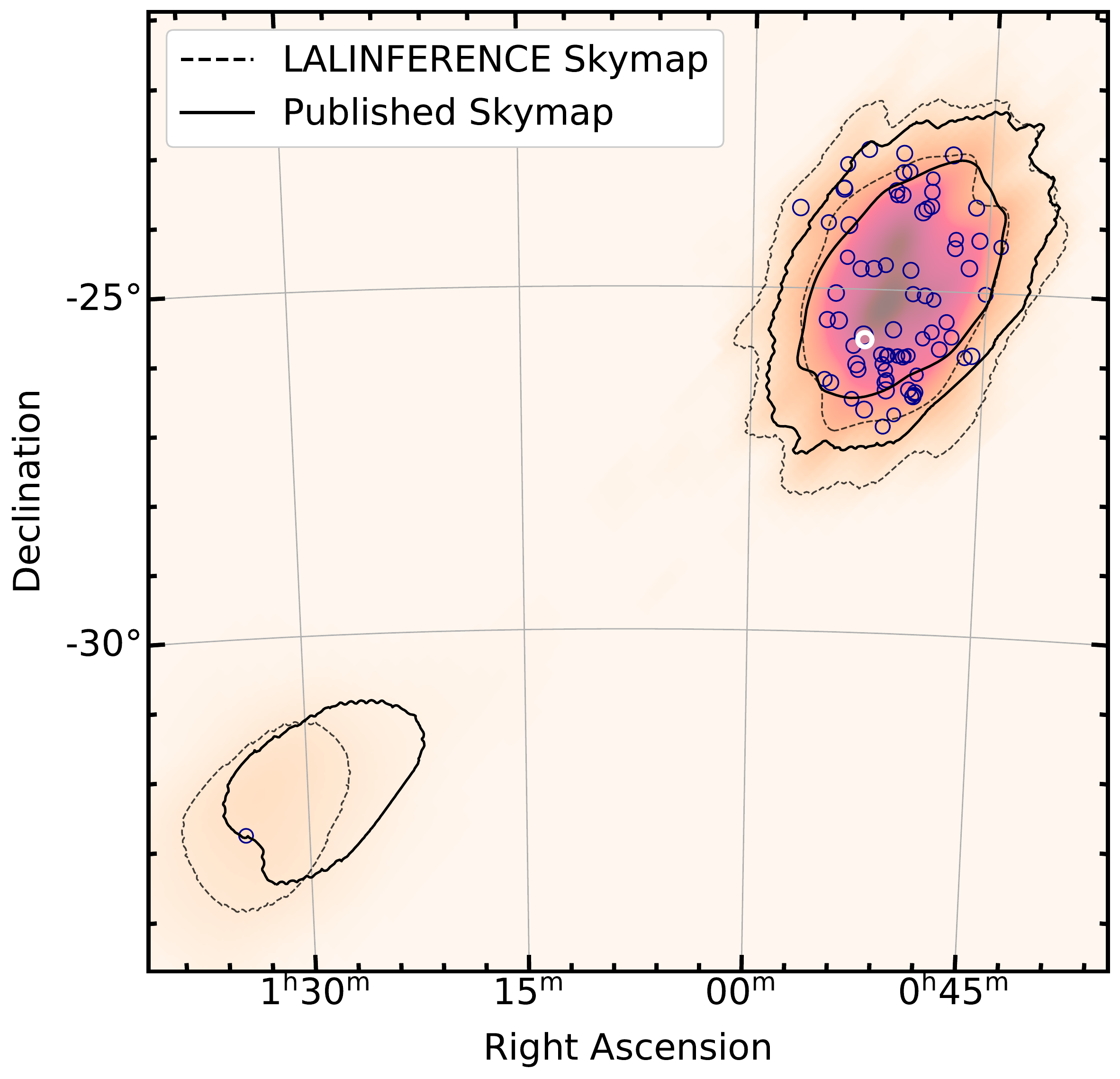}}}$
\hspace*{.1in}
$\vcenter{\hbox{\includegraphics[width=0.54\textwidth]{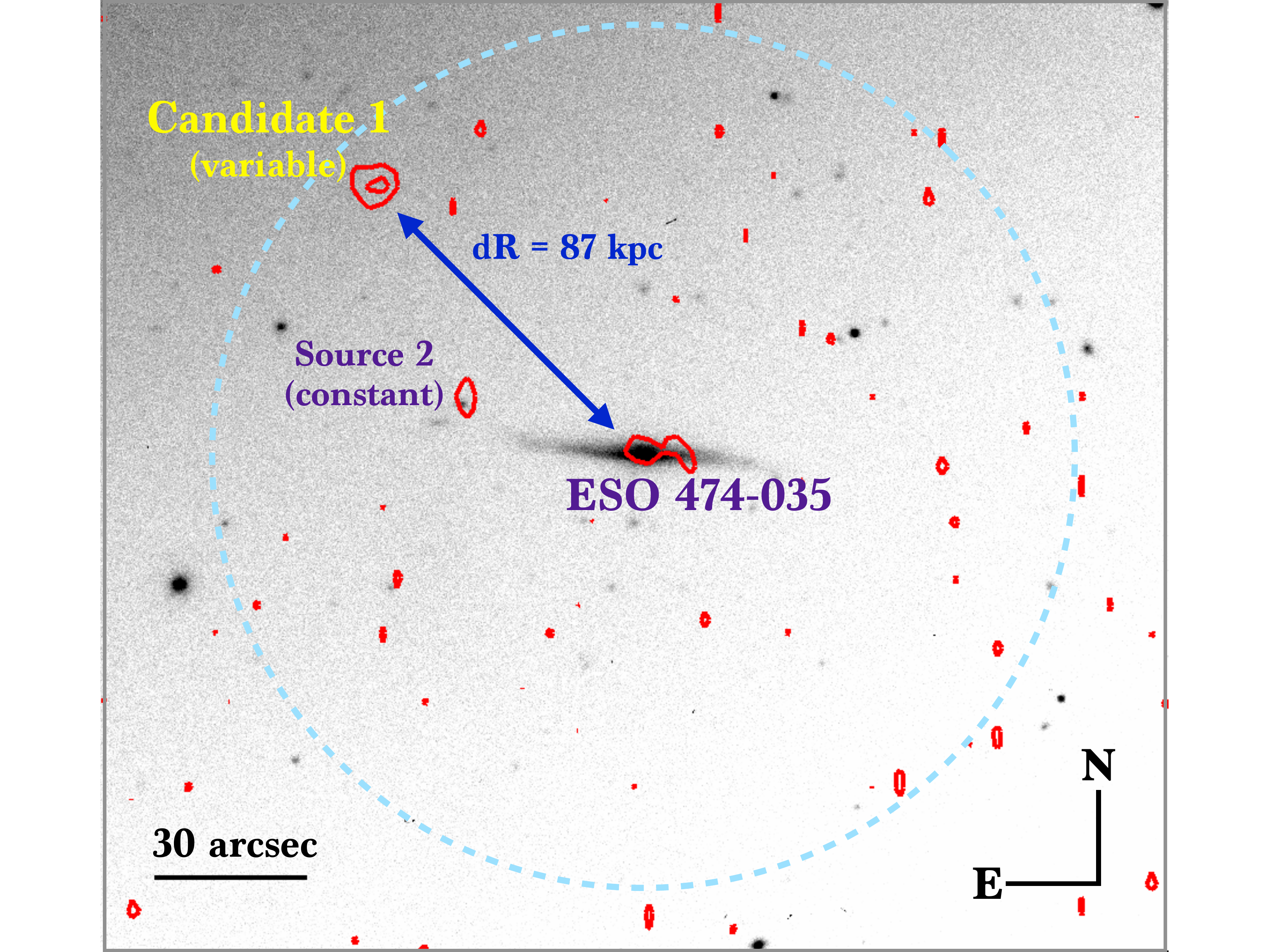}}}$
\caption{Left: The positions of our 75 galaxies (purple circles) in relation to the GW skymaps (dotted lines show the {\tt LALINFERENCE} map used to create our ranked list; solid lines show the revised final map from the full GW analysis, \citealt{lvc190814}). The size of each circle corresponds to the area searched around each galaxy (100 kpc physical offset, ${70-132}\arcsec$ depending on the galaxy distance). The position of the galaxy ESO 474-035 is highlighted in white. Right: The field containing ESO 474-035, as seen in the optical (grayscale, $i$-band image from Magellan taken 1.48~days post-merger) and the radio (red contours, 6 GHz image from our VLA program taken 213 days post-merger). We detect two significant point-like radio sources within our search region (dashed light blue line): ``Candidate 1" (variable, not detected at 35 days post-merger), and ``Source 2'' (consistent with constant flux density). We also observe weak extended emission near the nucleus of ESO 474-035. At the distance of ESO 474-035 ($d_L = 271$~Mpc), Candidate 1 would be at a projected offset of {87} kpc (blue line), at the upper end of the distribution measured from short GRBs. Unlike Source 2, which has a clear optical counterpart, Candidate 1 has no coincident optical emission to $i\gtrsim 22.23$~mag at 1.48~days after the merger, and no underlying, static host galaxy to $r = 24.92$~mag.}
\label{fig:magellan}
\end{figure*}

\subsection{Galaxy selection criteria}

To maximize the likelihood of observing the counterpart location with a minimal expenditure of telescope time, we selected a galaxy-targeted search strategy for GW190814. Galaxy-targeted searches are particularly appropriate for telescopes with smaller instantaneous fields of view (like the VLA). The galaxies we target include many of the most optically luminous galaxies within the localization volume, the same galaxies that have been prioritized for counterpart searches at optical and X-ray wavelengths. Such galaxies are attractive targets because they contain much of the stellar mass within the localization volume (and thus have the highest probability of actually containing the merger). However, they are also more likely to have detectable radio emission from unrelated sources (e.g.~star formation, AGN activity), making searches for radio transients more challenging. 

To select our targets, we generated a list of all galaxies from the Galaxy List for the Advanced Detector Era (GLADE) catalog \citep{glade} with $B$-band luminosities $L\gtrsim0.1L_*$ in the {\tt LALINFERENCE} 90\% localization volume that was circulated by the LIGO/Virgo collaboration 13.5 hr post-discovery \citep{gcn25333}. We then ranked this list based on a weighting of the galaxy's spatial position within the localization volume and the galaxy's $B$-band luminosity (a proxy for stellar mass), using the same procedure followed in \cite{Gomez2019} and \cite{Hosseinzadeh2019}. We observed the top 75 galaxies on this list, out of a total of 723 galaxies. The observed galaxies contained $\sim21$\% of the cataloged integrated stellar luminosity in the region. However, the GLADE catalog is known to be incomplete at the distance of GW190814 \citep{glade,lvc190814}. We therefore estimate the completeness of the catalog by integrating a Schechter B-band galaxy luminosity function down to $0.1L_*$ to approximate the true total number of galaxies and the corresponding total integrated stellar luminosity contained in the region (see \citealt{Gomez2019} for the exact function used). We find that the GLADE catalog is $\sim50$\% complete down to $0.1L_*$ at the distance of GW190814, in terms of number of galaxies. We estimate that our 75 galaxies contain 14\% of the total integrated stellar luminosity within the {\tt LALINFERENCE} localization volume.
 
The final localization volume presented in \cite{lvc190814} shifted slightly compared to the {\tt LALINFERENCE} localization volume (Figure \ref{fig:magellan}, left panel) and shrank significantly, from $1.1\times10^{5}$ Mpc$^3$ to $3.9\times10^4$ Mpc$^3$. We therefore repeat the above calculations for the final localization volume. We find that 65 of our target galaxies remain within the final 90\% localization volume presented in \cite{lvc190814}, comprising 32\% of the total integrated stellar luminosity within this region. Thus, assuming that the merger probability tracks stellar light, we have a roughly 32\% chance that our observations covered the true position of the merger.

\subsection{VLA observations}

We observed the top 75 galaxies in our ranked list in September 2019 ($\sim1$ month post-merger; ``epoch 1'') and again in February/March 2020 ($\sim6-7$ months post-merger; ``epoch 2''). To facilitate scheduling, we split each epoch into three 3h 20m scheduling blocks of 25 galaxies each, with the observing order chosen to minimize slew times. A complete list of the galaxies observed and the timing of each observation is given in Table \ref{tab:obs}. The position of each galaxy and the size of the region searched for transients in each pointing is shown in comparison to the GW localization in Figure \ref{fig:magellan} (left panel). The search region from a sample pointing is also shown in Figure \ref{fig:magellan} (right panel).

To maximize the chance of detection, we utilized the sensitive C band receiver in 3-bit mode (4 GHz bandwidth, mean observing frequency 6 GHz) for all observations. We used 3C147 as our flux and bandpass calibrator for all of our observations and one of three phase calibrators depending on which was closest to a given galaxy: J0011-2612, J0118-2141, or J0120-2701. We calibrated all data using the standard NRAO pipeline in CASA \citep{casa} and imaged the data using the CASA task {\tt clean}.

The first epoch of observations was taken when the VLA was in its most extended A configuration (beamsize $\sim0.33 \arcsec$ at 6 GHz) and the second epoch when it was in its more compact C configuration (beamsize $\sim3.5\arcsec$ at 6 GHz). As any physically-plausible radio afterglow at the distance implied by the GW signal ($d_L=241^{+41}_{-45}$ Mpc; \citealt{lvc190814}) is predicted to be an unresolved point source on the timescale of our observations, the differing resolution of the radio observations should not impact the flux density recovered for any (sufficiently isolated) bona fide counterpart. Nevertheless, the C configuration data is more sensitive to diffuse emission from the merger host galaxy or other sources in the field, which caused additional challenges with the data imaging for a subset of our targets. This manifests as an elevated rms noise level in a small fraction of our images. In epoch 1, we achieved a median image rms of 12.2 $\mu$Jy beam$^{-1}$ and in epoch 2, we achieved a median image rms of 17.6 $\mu$Jy beam$^{-1}$. In both epochs, our typical time on source was $\sim6$m 20s for each galaxy. 

\section{Counterpart Search}\label{sec:search}

\begin{figure*}
\centering
\includegraphics[width=0.95\textwidth]{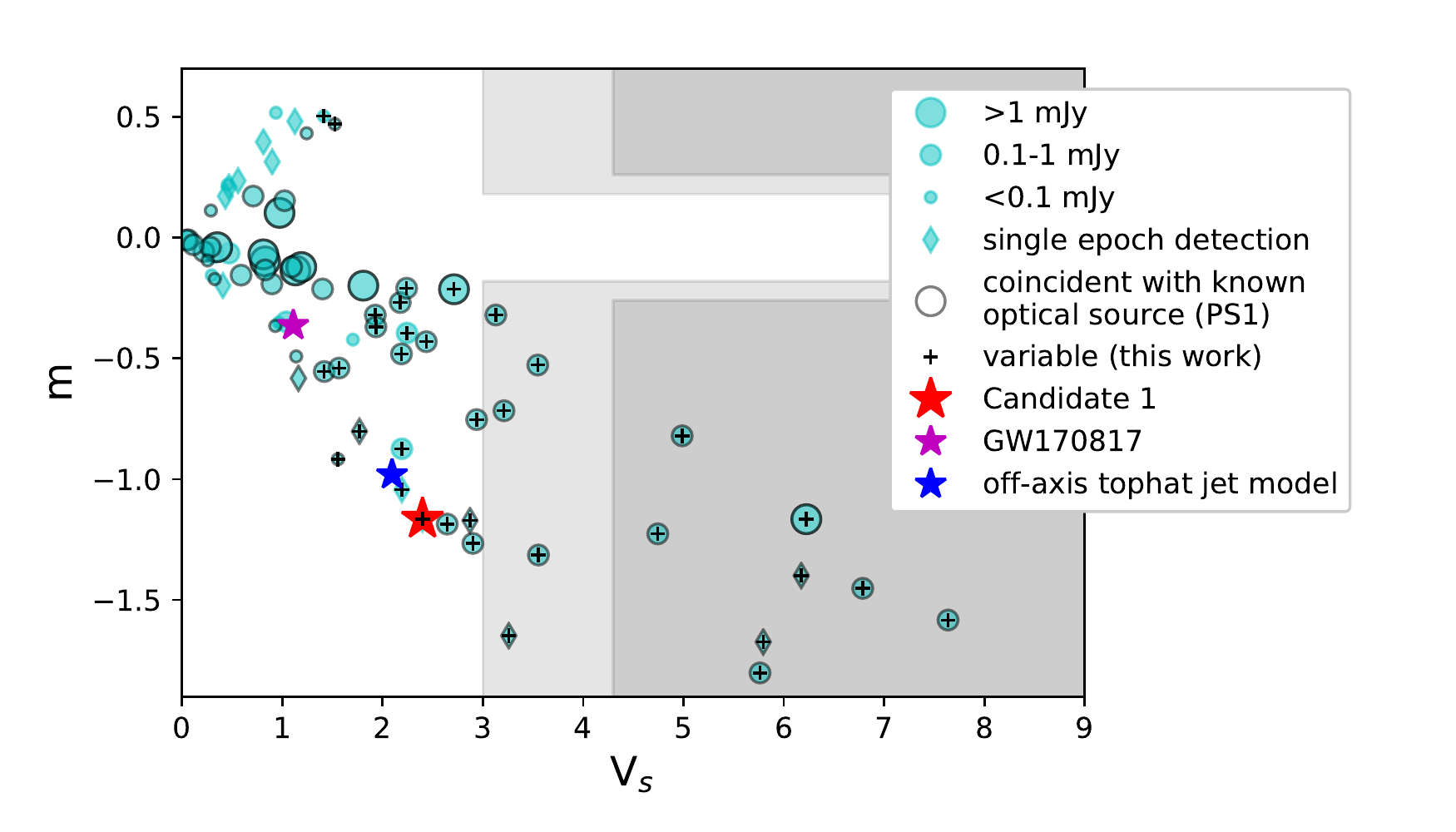}
\vspace{-0.2in}
\caption{The variability statistic ($V_s$) versus modulation index ($m$) for the population of radio sources detected in our observations (circles are sources detected in both epochs, diamonds are detected in only one epoch). The sources outlined in black are coincident with optical sources in archival data; we consider them less likely to be possible counterparts to GW190814. Sources within the darker gray shaded region would have been defined as significantly variable by \cite{Radcliffe19}, while the lighter gray region indicates the definition used by \cite{bhakta20}. However, these criteria may miss genuine GW counterparts: for example, GW170817's radio counterpart is not obviously highly variable on the timescale of our observations (magenta star). {Even jet emission that peaks at the time of our second epoch could be missed if observed with the cadence and sensitivity of our observations; one such model that we considered for GW190814 is shown by the blue star (a tophat jet with $E_{K,iso} =5\times10^{51}$ erg, n = 0.09 cm$^{-3}$, ${\epsilon_e=0.1, \epsilon_B=0.01, p=3.2,}$ and viewing angle $45^{\circ}$).} We utilize a less-restrictive variability definition in this work, resulting in a larger sample of possibly variable objects (black crosses). We select one of these variables for additional multi-frequency follow up, based on its large modulation index and its lack of an optical counterpart (Candidate 1, red star).}
\label{fig:variability}
\end{figure*}

\subsection{Candidate Selection Criteria}\label{sec:criteria}

We next searched each galaxy field for a radio counterpart to GW190814. To identify radio sources, we used Source Extractor \citep{se} in combination with the distance from the GLADE catalog\footnote{Distances are calculated using an assumed flat $\Lambda$CDM cosmology with $H_0=70$ km s$^{-1}$ Mpc$^{-1}$, $\Omega_M=0.27$, and $\Omega_{\Lambda}=0.73$.} to locate all radio sources detected with $>5\sigma$ significance in each image located within 100 kpc of each of our target galaxies. We chose this search radius because $\gtrsim 95\%$ of observed SGRBs are found within 100 kpc of their hosts \citep{fong13}; thus, if GW190814 belongs to the population of mergers that produce SGRBs then we expect that a real radio counterpart to GW190814 will likely be found within this area. This resulted in an initial list of {72} detected sources in epoch 1 and {102} detected sources in epoch 2. We then visually inspected all images to remove sources that were obvious imaging artifacts or clearly extended, leaving {75} sources detected in at least one epoch. We measured the flux densities of each detected source using the {\tt imtool fitsrc} command within the {\tt pwkit} package \citep{pwkit}. All flux densities were extracted assuming a point source fit. We found that due to the differing rms noise level in the epoch 1 and 2 images, some sources that were only recovered by Source Extractor with $>5\sigma$ significance in one of the two epochs may nevertheless be consistent with constant flux density. For each source detected in only one epoch, we therefore ran {\tt fitsrc} at the source position in the other epoch to search for lower-significance emission. We then used the recovered flux density (or 3 times the image rms at the source position if no emission was detected) to measure the variability of each object. 

We assess the variability of our radio sources in several ways. Several previous wide-field searches for radio transients have characterized variability between two epochs in terms of the modulation index, $m$, and the variability t-statistic, $V_s$ (e.g.~\citealt{Mooley16,Radcliffe19,bhakta20}). We define:

\begin{equation}
    m = \frac{\Delta S}{\langle S\rangle} = 2\frac{S_1-S_2}{S_1+S_2}
\end{equation}
where $S$ is the flux density of each source, as determined from a point source fit with {\tt fitsrc}, and
\begin{equation}
    V_s = \left|\frac{\Delta S}{\sigma}\right| = \left|\frac{S_1-S_2}{\sqrt{\sigma_1^2+\sigma_2^2}}\right|
\end{equation}
where $\sigma$ is the total measurement error on the flux density. We include both the uncertainty derived from the point source fit and an additional error term of 5\% corresponding to the accuracy of the absolute flux calibration scale of the VLA \citep{pb17} in this quantity. We plot the distribution of our radio sources in $V_s$ and $m$ in Figure \ref{fig:variability}. 

Previous work has often focused on maximizing the purity of constructed samples of radio variables and transients, and thus has imposed fairly strict cutoffs for variability: \cite{Mooley16} and \cite{Radcliffe19} require $V_s>4.3$ and $|m|>0.26$ (darker gray shaded region in Figure \ref{fig:variability}), while \cite{bhakta20} require $V_s>3$ and $|m|>0.18$ (light gray shaded region).\footnote{This is equivalent to demonstrating variability at the $>4\sigma$ or $>3\sigma$ confidence level, respectively, in the case of Gaussian noise. More generally, $V_s\leq4.3$ is the 95\% confidence interval for the $t$-statistic.} Applying these criteria, we find 8 and 13 variable sources respectively in our sample, corresponding to a variability fraction of {11}\% and {17}\%. This is significantly higher than previous blind untargeted searches, which have found that only a few percent of radio sources over large sky areas exhibit this level of variability (e.g.~\citealt{Carilli+03,frail12,Mooley16,bbm+18,Radcliffe19,Dobie2019,bhakta20,sarb20}). We note that all 13 variables are coincident with bright galaxies in archival optical imaging, and 12 of them are coincident with the nuclei of our target galaxies to within astrometric errors. One natural explanation for the higher prevalence of variability in our sample compared to previous work is thus that the centers of (relatively) nearby galaxies selected as likely hosts for GW mergers are more likely than average regions of space to contain sources of detectable radio emission (e.g.~weak AGN). We explore this further in Section \ref{sec:contamination}. 

One downside of optimizing for sample purity in our data is that GW radio counterparts are expected over a broad range of timescales: weeks to months for relativistic jet afterglows, or months to years or even decades for kilonova afterglows (e.g.~\citealt{np11}). Thus, with only two epochs of data, it is worth exploring additional, conservative methods to ensure that transients with variability timescales not well-aligned to our observing cadence are also discovered. For example, such techniques would be necessary to recover a GW170817-like radio transient in our data: if we compute $m$ and $V_s$ for GW170817 using the 6 GHz observations collected closest to the timescale of our observations (at 39 d and 217 d respectively; \citealt{alex17,alex18}), it would not be classified as a variable source by the methods outlined in either \cite{Radcliffe19} or \cite{bhakta20} (Figure \ref{fig:variability}, magenta star). Even an off-axis relativistic jet that peaks on the timescale of our second epoch may not satisfy the criterion for $V_s$ given the sensitivities achieved in our two epochs (e.g.~sample model in Figure \ref{fig:variability}, blue star).

We are thus motivated to explore a less stringent criterion for variability, to emphasize completeness of our sample rather than purity. We therefore create a list of all radio sources inconsistent with constant flux density to within the $>1\sigma$ measurement uncertainties calculated by {\tt fitsrc} (using 3 times the image rms at the source position as an upper limit on the flux density for sources only detected in one epoch). We identified {34} potential radio counterparts using this criterion (Figure \ref{fig:variability}, black crosses). {32} of our {34} variable sources (including all 13 of the ``highly significant variables" satisfying the \citealt{bhakta20} criteria discussed above) increased in brightness between epochs 1 and 2, while only two sources decreased in brightness. We do not expect this imbalance to result from differences in absolute flux calibration between the two epochs, as the variables were not preferentially observed in any single scheduling block (and as mentioned above, we already include a conservative additional 5\% uncertainty term in the measurement errors used to compute $V_s$, based on the known accuracy of the VLA flux calibration scale). Instead, the apparent increase in flux density of many of our sources may be partially explained by the reconfiguration of the VLA between epochs 1 and 2: the C configuration epoch 2 data are more sensitive to diffuse low surface brightness emission (from e.g.~ongoing star formation) than the A configuration epoch 1 data. (This is also consistent with the fact that a number of our radio sources appear point-like in epoch 1 and slightly extended in epoch 2, despite the epoch 2 data having $\sim10\times$ lower resolution.) 

The centers of galaxies may be particularly prone to such spurious detections of variability, as they may contain a superposition of extended and compact emission components from star formation and/or AGN activity. Indeed, the positions of 17 of our variable sources are consistent with the centers of their target GLADE galaxies to within astrometric uncertainties. While some models predict an enhanced rate of compact object mergers near the supermassive black holes at the centers of galaxies \citep{ap12,mck20,Perna21,Zhu21}, this result suggests that GW counterparts near galactic nuclei will be particularly challenging to identify in the radio. 
We rule out these 17 sources as likely radio counterparts to GW190814.

\begin{figure*}
\centering
\includegraphics[width=\textwidth]{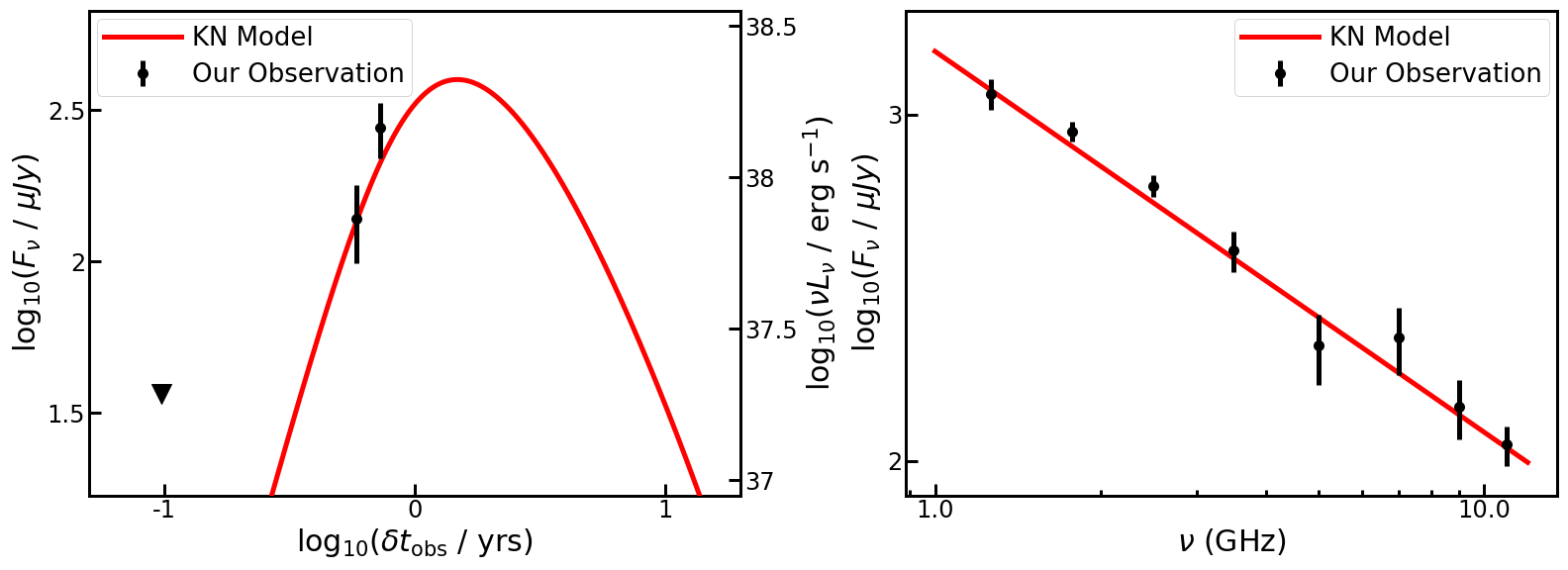}
\vspace{-0.2in}
\caption{The 6.0~GHz light curve (left) and spectral energy distribution at 266 d (right) of Candidate 1 (black points, error bars are $1\sigma$). The triangle indicates a $3\sigma$ upper limit.
The red lines represent one allowed model from our kilonova ejecta modeling ($n = 10^{-4}$~cm$^{-3}$, $E_{\rm ej}=3.6 \times 10^{52}$~erg, $M_{\rm ej}=0.01 M_\odot$, $p=3.2$, $\epsilon_e = 0.1$, $\epsilon_B = 0.01$).
}
\label{fig:02WGO}
\end{figure*}

Finally, we searched the VLA Sky Survey (VLASS; \citealt{vlass}) Quick Look images and the PanSTARRS-1 data archive (PS1; \citealt{chambers16}) at the positions of the remaining candidates to provide additional insight into their nature. {Three} sources are detected in VLASS epoch 1 data that pre-dates GW190814, suggesting that they are unrelated to the merger. Furthermore, we found that all but six sources had a spatially coincident optical counterpart in the PS1 catalog, suggesting that these radio sources are also likely not plausible counterparts to GW190814. While the exact peak timescale of a hypothetical radio counterpart to GW190814 is uncertain, models of off-axis relativistic jets with $\theta_{obs}\sim46^{\circ}$ (GW190814's binary inclination) and parameters typical of short GRBs are expected to show significant variability between the timescales of our two epochs (Figure \ref{fig:variability}, blue star). Of our remaining six candidates, the source that shows the largest $|m|$ is located at R.A. = \ra{00}{52} {45}.418, Dec = \dec{-25}{43}{08}.16 (J2000), within the field of the galaxy ESO 474-035, which was ranked sixth in our catalog of 75 galaxies (Figure \ref{fig:magellan}). We discuss this source (hereafter ``Candidate 1") and its evolution in more detail in the next section.

\subsection{Modeling of Candidate 1 in the Field of ESO 474-035}

We now focus on Candidate 1 to assess its viability as a counterpart to GW190814. Previous Magellan observations at 1.48~d after the GW trigger cover both the catalogued GLADE galaxy ESO 474-035 and the position of Candidate 1, and placed a limit of $i>22.23$~mag on any transient optical emission at that time (Figure \ref{fig:magellan}; see also \citealt{Gomez2019}). Moreover, deeper, pre-merger limits at the candidate position from the Legacy Survey \citep{Dey2019} of $r = 24.92$~mag, derived from the 5$\sigma$ depth of co-added DECam images (Brick ID 0131m257), place stringent constraints on an underlying source. For a constraint on a satellite galaxy at the same distance as ESO 474-035, this translates to a $L\lesssim 4.1 \times 10^{6}\,L_{\odot}$. This is roughly 4 orders of magnitude below the luminosity of the Milky Way, ruling out all except the faintest dwarf galaxy regime \citep{Simon19}. Moreover, the lowest host galaxy luminosities derived for short GRBs are $\approx 10^{9}\,L_{\odot}$ \citep{ber14}, well above the limit derived here. If instead there is a background galaxy at a similar luminosity to the Milky Way at the position, it would need to be at $z \gtrsim 2$ to be consistent with this limit, implying a luminosity of $\gtrsim3\times10^{41}$ erg s$^{-1}$ for the radio transient, comparable to radio-loud quasars \citep{Kellerman16}. Thus, we find that while a background quasar cannot be ruled out, if the origin of Candidate 1 is from a stellar progenitor, then it likely originated from ESO 474-035. 

We triggered additional multi-frequency radio observations of Candidate 1 at $1-12$ GHz, which were carried out on 2020 May 5 ($t=266$ d, ``epoch 3"), with the VLA in C configuration. Our observations reveal a continued increase in the flux density at 6 GHz (Figure \ref{fig:02WGO}, left), confirming that the change in flux density is intrinsic to the source, rather than an artifact of the VLA configuration change between epochs 1 and 2. The broadband spectrum is optically-thin, consistent with a single power law $F_{\nu}\propto\nu^{\beta}$, where $\beta = -1.1 \pm 0.06$ (Figure \ref{fig:02WGO}, right). At the highest frequencies observed ($8-12$ GHz), we note that the emission appears partially resolved into two components, with centroids separated by $\sim3\arcsec$. At the distance of ESO 474-035 (271 Mpc) this would correspond to a physical separation of $\sim4$ kpc, orders of magnitude too large for a newly-formed GRB jet. This may suggest that our candidate is instead a background double-lobed radio AGN jet undergoing a flaring event. 

AGN are one of the most common types of compact sources in the radio sky, and in the radio their flares have typical timescales of months to a few years (e.g.~\citealt{hov08}). Thus, AGN flares are likely to be the largest source of contamination in searches for radio GW counterparts. We note that prior to obtaining data with better spatial resolution in epoch 3, Candidate 1 initially appears very similar to expectations for a GW counterpart, and we may expect to find similar contaminants in future GW counterpart searches where high-resolution data are not available. We therefore invest some additional effort in fitting Candidate 1's radio emission with models appropriate for GW counterparts, to see if we can distinguish Candidate 1 from a typical radio GW counterpart based on the physical properties required to fit the flux density alone. For this analysis, we use the combined flux density of the two resolved components when modeling the high-frequency epoch 3 data and we assume that Candidate 1 is at the distance of ESO 474-035. 

We consider two classes of radio GW counterpart models for Candidate 1: radio emission from collimated fast ejecta (i.e. an initially relativistic jet, possibly with some velocity structure) and from slower ejecta (the same material in which r-process nucleosynthesis occurs at early times, producing the kilonova optical transient; assumed to be quasi-spherical and moving at up to a few tens of percent of $c$). In both cases, the radio emission is synchrotron radiation arising from a population of electrons accelerated into a power-law distribution of energies, $N(\gamma) \propto \gamma^{-p}$ for $\gamma>\gamma_m$, as the merger ejecta shocks and interacts with the surrounding interstellar medium. The allowed parameter space is highly degenerate, as we observe only the rising portion of the light curve and a single power-law segment of the synchrotron spectral energy distribution. We find $p = 3.2 \pm 0.11$ using the spectral slope computed from our multi-frequency observations ($\beta = -1.1 \pm 0.06$), assuming the radio observations lie above $\nu_m$ (the synchrotron frequency corresponding to $\gamma_m$) and below the cooling frequency ($\nu_c$) \citep{gs02}. This value is in line with some supernovae (see e.g. \citealt{vandyk1994, Chevalier98}), although it disagrees with the precise value $p=2.15^{+0.01}_{-0.02}$ calculated for GW170817 \citep{Hajela19} and it is higher than previously observed in many cosmological SGRB afterglows, where typically $2<p<3$ \citep{fong15}.

We first consider the possibility that Candidate 1's radio emission is due to an off-axis relativistic jet. We fix $p=3.2$ and $\theta_{obs}=46^{\circ}$ (to put the jet in alignment with the best-fit binary inclination as derived from the GW signal; \citealt{lvc190814}), and we assume that the fractions of energy carried by electrons and magnetic fields in the shock are $\epsilon_e = 0.1$ and $\epsilon_B = 0.01$, respectively. While this agrees with assumptions made in previous studies of cosmological SGRBs, e.g.~\cite{fong15} {and with previous theoretical and observational studies that find $\epsilon_e\approx0.1$ for relativistic shocks (e.g.~\citealt{Spitkovsky08,ben17})}, we note that for most events at cosmological distances $p, \epsilon_e$ and $\epsilon_B$ are poorly constrained {by the data directly} and must be assumed. The full GW170817 dataset suggests that $\epsilon_B$ may be much lower than 0.01 in at least some jets \citep{mc20rev}; if this is true for GW190814, then the constraints derived below are also affected. {For example, models with lower values of $\epsilon_e$ or $\epsilon_B$ will have a larger total energy.} We consider both tophat jets (in which all of the jet energy is contained within a narrow cone with opening angle $15^{\circ}$) and jets with a relativistic core surrounded by Gaussian wings of slower-moving material. We find that only the tophat jet models can reproduce the steep rise seen in Candidate 1's 6 GHz light curve, and furthermore that we require a large isotropic-equivalent jet energy ($E_{K,iso} \sim8\times10^{53}$ erg) and a high density ($n \sim0.5$ cm$^{-3}$); comparable to the largest energy values and in the top 40\% of density values inferred for SGRBs \citep{fong15}. While we cannot entirely rule out the possibility that Candidate 1 is an off-axis relativistic jet launched by GW190814, we disfavor this possibility due to the large energy required and the high value of $p$. The high density would also be unexpected for such a highly-offset transient, particularly due to the lack of any optical emission at the transient position to suggest a satellite galaxy or globular cluster environment for the transient.

\begin{figure}
\centerline{\includegraphics[width=0.5\textwidth]{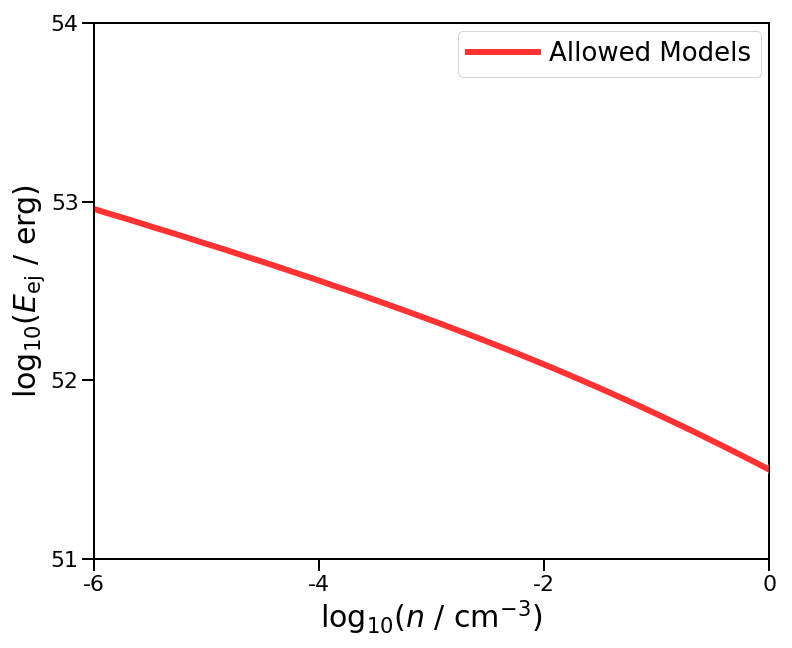}}
\caption{The allowed $E_{\rm ej}$ vs $n$ phase space for Candidate 1, under the assumption that the radio emission is produced by the shock between $0.01M_{\odot}$ of quasi-spherical ejecta and the ambient medium (with $p=3.2$, $\epsilon_e = 0.1$, and $\epsilon_B = 0.01$). Densities higher than 1 cm$^{-3}$ are ruled out by the 6 GHz light curve.}
\label{fig:02WGO_Evsn}
\end{figure}

We next compare the radio behavior to slow kilonova (KN) ejecta models to determine whether they are consistent with our observations. Following the prescriptions of \cite{Schroeder2020}, we modeled the 6 GHz lightcurve and multi-frequency spectrum of Candidate 1 with a KN ejecta interaction model. We again fix $p=3.2$, $\epsilon_e = 0.1$, and $\epsilon_B = 0.01$ in our modeling. Optical and near-infrared follow-up studies of GW190814 have placed constraints on the ejecta mass of $M_{\rm ej} \lesssim 0.04-0.1~ M_\odot$ \citep{Gomez2019, Kawaguchi2020, Andreoni2020, Morgan2020, Ackley2020, Thakur20, Viera20}. We set the ejecta mass $M_{\rm ej}$ to $0.01 M_\odot$, as lower values of $M_\odot$ would cause the time of observation of epoch 3 to approach the deceleration time, $t_{\rm dec}$, of the ejecta. 
If $t_{\rm dec} \approx t_{\rm obs,3}$, the time of observation of epoch 3, the KN model light curve would start to decline, whereas we observe the 6 GHz light curve still rising through epoch 3. 

\begin{figure}
\centering
\includegraphics[width=0.48
\textwidth]{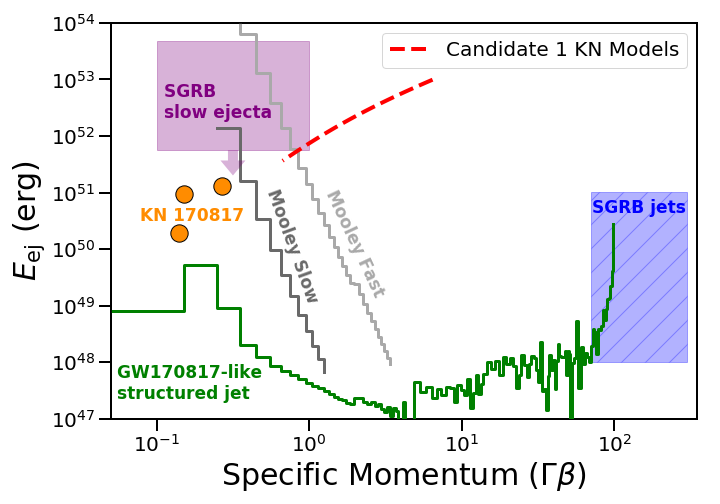}
\vspace{-0.2in}
\caption{The $E_{\rm ej}$ vs Specific Momentum ($\Gamma \beta$) phase space for KN models consistent with Candidate 1's radio evolution (red dashed line) in comparison to other ejecta models proposed for compact object mergers. Both quasi-spherical KN ejecta models and relativistic ejecta models struggle to reproduce the high energy required to match Candidate 1's radio properties. The orange circles show the energy of the red, blue, and purple KN components associated with GW170817 from \citealt{Villar17}, while the gray lines show two different models for the velocity distribution of quasi-spherical ejecta in this event \citep{mool18}. The purple shaded region is a representative range of maximum energies found for SGRB slow ejecta derived from late-time radio observations \citep{Schroeder2020}. The green line is the structured jet model for GW170817 from \cite{marg18}, while the blue shaded region is the beaming-corrected energy of the jet component in SGRBs \citep{fong15}. } 
\label{fig:EejVej}
\end{figure}

Even after making these assumptions, several parameter degeneracies remain in our modeling. We therefore create a grid of 250 light-curve models exploring a range of values for the density $n$ and ejecta energy $E_{\rm ej}$. We find a broad range of combinations that are consistent with our observations at the times of the two 6 GHz detections (Figure \ref{fig:02WGO_Evsn}, solid line). Densities above $n\sim1$ cm$^{-3}$ are ruled out, as these models would also require the light curve to begin declining by the time of our epoch 3 observation. As with the relativistic jet models, the ejecta energy required to match the observations is large, particularly for the lower densities expected given the large offset of Candidate 1. These models also require very high ejecta velocities, $\beta_{0} \sim 0.80-0.94~c$ for $n\sim0.01-10^{-4}$ cm$^{-3}$ (Figure \ref{fig:EejVej}, red dashed line). While the distributions of ejecta velocities in some published KN models have tails to high velocities $\sim0.9c$ (e.g.~Figure \ref{fig:EejVej}, gray lines), the bulk of the energy must be carried by slower-moving ejecta ($\approx 0.1-0.5~c$), to match the optical and infrared properties of the KN emission \citep{Bauswein13, Hotokezaka13, Sekiguchi16, Ciolfi17, Sekiguchi16, mool18}. {The ejecta energy is only mildly sensitive to the assumed values of $\epsilon_e$ and $\epsilon_B$; if energy equipartition is assumed ($\epsilon_e=\epsilon_B=0.33$) the required $E_{ej}$ decreases by a factor of $2-3$ across the range of allowed densities. Conversely, models with lower values of $\epsilon_e$ or $\epsilon_B$ require a higher $E_{ej}$.} It is therefore difficult to explain the high ejecta energy required to fit our radio light curve with existing KN models. 

In summary, both relativistic jet models and quasi-spherical KN ejecta models ultimately require very high energies and fast ejecta velocities to match the radio evolution of Candidate 1. A comparison to ejecta models consistent with GW170817 and cosmological SGRBs can be seen in Figure \ref{fig:EejVej}. It is clear that either the KN models used to explain Candidate 1 are probing a new regime of parameter space, one with higher energies and velocities than the other models that have been found to be consistent with the population of cosmological SGRBs and GW170817, or that some parameters (e.g.~$\epsilon_e$ and $\epsilon_B$) differ from the standard values we assumed. Nevertheless, even for higher $\epsilon_B$ values it is difficult to construct a plausible physical scenario that accelerates sufficient ejecta to such high velocities, particularly for compact object mergers with highly unequal mass ratios like GW190814. Ultimately, we conclude that this analysis disfavors Candidate 1 as a radio counterpart to GW190814 on physical grounds. We suggest that similar analyses can be applied in future GW counterpart searches to discriminate between genuine radio GW counterparts and unrelated background AGN.

\section{Discussion}\label{sec:disc}

\begin{figure*}
\centering
\includegraphics[width=0.49\textwidth]{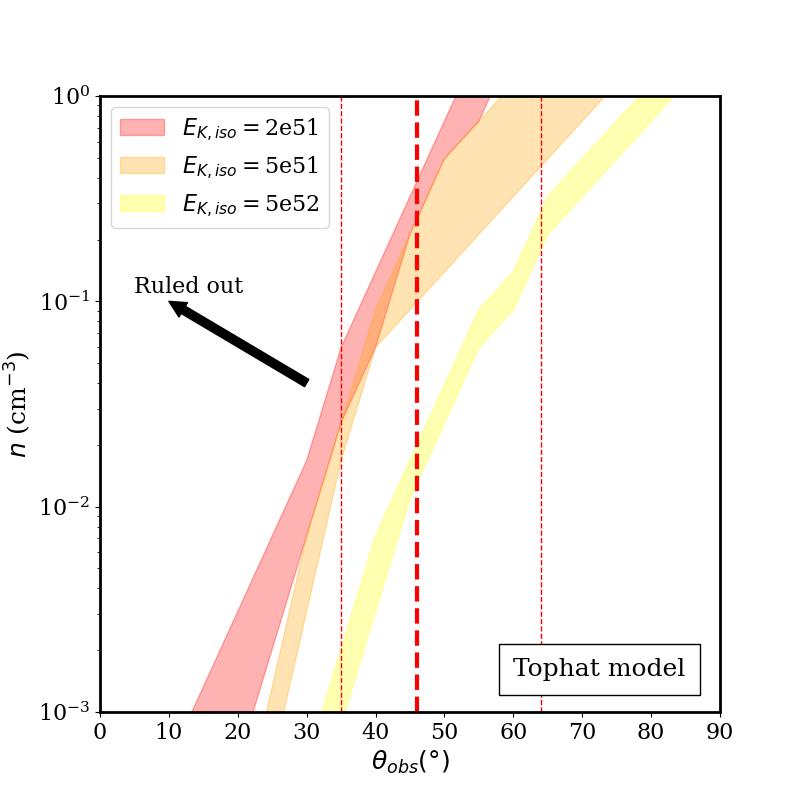}
\includegraphics[width=0.49\textwidth]{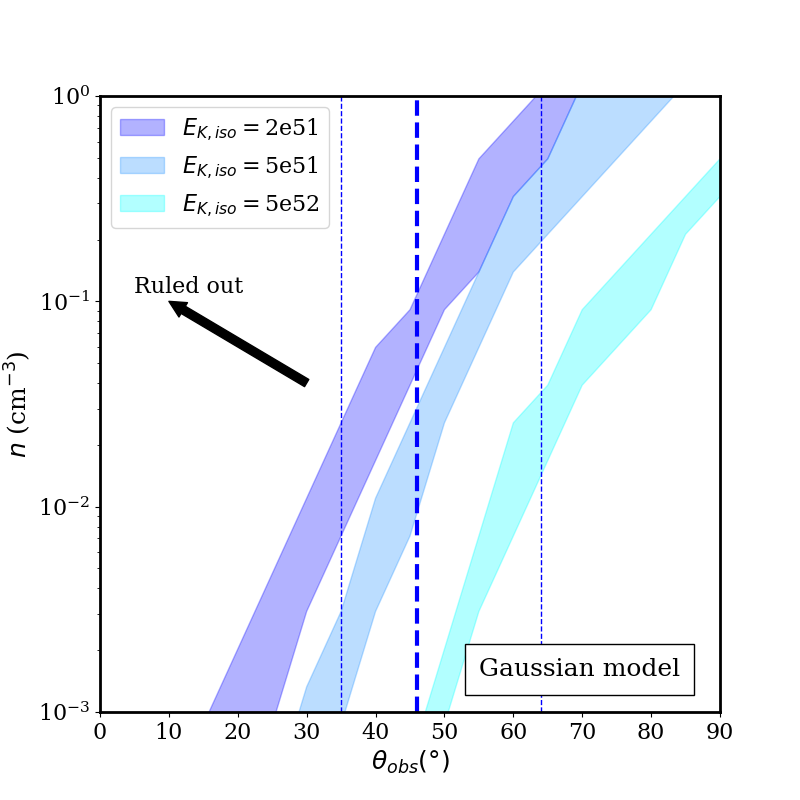}
\caption{Portions of parameter space ruled out by the non-detection in our two epochs at $\sim38$ and $\sim208$ days for a 15$^{\circ}$ tophat (left) and Gaussian (right) jet, with $E_{K,iso}$ = (0.2, 0.5, and 5)$\times10^{52}$ erg, $p=2.2, \epsilon_e=0.1$, and $\epsilon_B=0.01$. The shaded region shows the uncertainty from the distance. All space above and to the left of these limits is ruled out by our non-detections. Under these assumptions, we can rule out an energetic jet similar to those seen in cosmological short GRBs for a range of viewing angles and densities. The binary inclination angle and its associated uncertainty calculated from the full GW analysis \citep{lvc190814} are shown by the dashed lines.}
\label{fig:02WGO_jets}
\end{figure*}

\subsection{Limits on a Relativistic Jet Launched by GW190814}\label{sec:jets}

Apart from Candidate 1, we detect no other convincing radio counterpart to GW190814 in our observations. We therefore next investigate the limits we can place on the existence of a relativistic jet (assuming that the merger occurred within one of the galaxies we targeted). We generate a grid of light curves at 6 GHz using {\tt afterglowpy} \citep{afterglowpy} for 2 relativistic jets: 1) a tophat jet with an opening angle $\theta_{jet} = 15^{\circ}$, and 2) a Gaussian jet with a $15^{\circ}$ jet core ($\theta_{jet}$) and wings extending out to $6~\theta_{jet}$. {Similar jet opening angles are seen in many GRBs \citep{Ryan15}.} We compute the light curves for isotropic-equivalent kinetic energies, $E_{K,iso}$ = $2\times10^{51}$, $5\times10^{51}$, and $5\times10^{52}$ ergs (representing a typical SGRB energy, as well as two more optimistic cases), over $0.1-1000$ days. These light curves are computed on a fine density grid with varying circumburst density $n = 10^{-6} - 10^{3}$ cm$^{-3}$, and for viewing angles ranging for 0$^{\circ}$ (on-axis) to 90$^{\circ}$. We fix $p$ = 2.2, $\epsilon_e=0.1$, $\epsilon_B=0.01$, and $d_L=241^{+41}_{-45}$ Mpc. We then compare these light curves to the limits placed by our VLA observations at 38 and 208 days, using {5}$\times$the typical RMS of our images in each epoch as an upper limit on the flux density of any counterpart at that time (Figure \ref{fig:02WGO_jets}). These limits allow us to rule out the higher density parameter space, as higher densities result in a brighter source which would have been detected in our observations. Alternately, if the value of $\epsilon_B$ is significantly lower than we assumed, then higher density models may still be allowed by our data.

Compared to the best-fit viewing angle of $\theta_{\rm obs}=46^{\circ}$ calculated by LIGO/Virgo, we find our observations rule out densities ${n \gtrsim}$ {0.4, 0.3, 0.02} cm$^{-3}$ assuming a top-hat jet model for $E_{K,iso}$ = $2\times10^{51}$, $5\times10^{51}$, and $5\times10^{52}$~erg respectively. Compared to the circumburst densities found for the SGRB population (assuming $\epsilon_e$ = 0.1 and $\epsilon_B$ = 0.01 for consistency; \citealt{fong15}), these constraints are comparable to the higher end of the population (at the ${\sim60-85\%}$ 
level compared to SGRB densities). These limits are a strong function of the viewing angle; for the jet with $E_{K,iso}$ = $2\times10^{51}$ erg, the lower limit on the density ranges from ${0.06-4}$ 
cm$^{-3}$ for the full range of viewing angles $35^{\circ}-64^{\circ}$ allowed by the GW analysis. For Gaussian jet models with the same energies, we find more constraining densities of ${n \gtrsim 0.1}$, {0.03}, ${9 \times 10^{-4}}$ 
cm$^{-3}$ (at the ${\sim40-65\%}$ 
level of SGRBs) for $\theta_{\rm obs}=46^{\circ}$ (${n\gtrsim0.03-1}$ cm$^{-3}$ 
for {the ${2\times10^{51}}$ erg jet observed at} $\theta_{\rm obs}=35^{\circ}-64^{\circ}$ at $d_L=241$ Mpc). {Thus, the assumed jet structure (tophat versus Gaussian) can affect the implied limits on circumburst density by a factor of a few. Our limits additionally become less constraining for more narrowly-beamed jets (which peak earlier); we can rule out a tophat jet with $E_{K,iso}$ = $2\times10^{51}$ erg and $\theta_{jet}=10^{\circ}$ viewed at $\theta_{\rm obs}=46^{\circ}$ for densities $n>1$ cm$^{-3}$.} In comparison, the search conducted by \cite{Dobie2019} can rule out comparable densities for a {tophat} jet with $E_{K,iso}=10^{51}$ erg, $\theta_{jet} = 10^{\circ}$, $\epsilon_e=0.1$, $\epsilon_B=0.01$, and $p=2.2$ only if the jet is viewed at $\lesssim{35}^{\circ}$ off-axis; for a jet that is $46^{\circ}$ off-axis, the density is not constrained by their data. This emphasizes the importance of continuing radio transient searches to late times ($\gtrsim6$ months post-merger) to fully constrain the allowed parameter space for highly off-axis jets, matched to their later peak timescales.

\subsection{Properties of our most variable sources: extreme AGN flares?}

The differing resolution and sensitivity of our data may impact the interpretation of sources with both small and large apparent flux density changes on the timescale of our observations, as discussed in Section \ref{sec:criteria}. Nevertheless, five of of our 13 ``highly-variable sources" (selected using the definition of \citealt{bhakta20}) have strong detections in VLASS data predating 2019 August 18 and three have marginal detections, confirming that they are likely unrelated to GW190814. The brightest of these radio sources is coincident with ESO 474-026 (the most highly-ranked galaxy on our prioritized list, at a distance of 244 Mpc), and appears strongly point-like in both our observations and the VLASS quicklook data. We therefore suggest that its variability is intrinsic, and that this radio source is a compact AGN. Interestingly, the flux density of this source increases dramatically, from $0.34\pm0.03$ mJy in epoch 1 to $1.31\pm0.14$ mJy in epoch 2. This factor $\sim4$ increase in flux density over a timescale of $\sim5$ months is not unprecedented for AGN flares, although their typical fractional variability is much lower at 6 GHz (from a few percent to a factor of $\lesssim2$; \citealt{hov08}). {Additionally, \cite{cs04} reported a 6 GHz flux density of $2.5\pm0.1$ mJy for this source in VLA observations taken on 1994 October 2 at similar resolution ($5\arcsec$), suggesting that its large-amplitude radio variability is not unique to the period of our observations. ESO 474-026 has been classified as a polar ring galaxy; its unusual optical morphology is most likely a short-lived state produced by a recent major galaxy merger \citep{resh+05,Spavone+12}. We suggest that this merger event could have fueled the AGN responsible for the variable radio emission.} 

Some models predict an enhanced rate of BNS and NSBH mergers in the accretion disks surrounding AGN \citep{mck20,Perna21,Zhu21}, but the resulting radio transient would be difficult to disentangle from the type of variability seen in ESO 474-026. Our results suggest that mergers in active galaxies will be particularly hard to discover and model in radio-only datasets, as radio AGN variability is still poorly constrained on the relevant timescales. This highlights the importance of continued monitoring of such sources, given the wide range of timescales expected for extragalactic radio transients. Ongoing and planned all-sky radio surveys are beginning to provide this vital long-term coverage; this particular source will next be observed in February 2022 during epoch 2.2 of VLASS.

\subsection{Considerations for Future Galaxy-Targeted Radio Counterpart Searches}\label{sec:contamination}

Finally, we briefly expand upon some implications of the high variability fraction of our sample of radio sources. Compared to untargeted searches, nearby galaxies are likely to have a higher surface density of detectable radio transients with unresolved, compact emission, including radio supernovae, tidal disruption events (TDEs), and AGN flares {(e.g.~\citealt{Weiler+86,Weiler+02,Berger+03,hov08,Alexander20};} \citealt{rc11, metzger15, Alexander15, Irwin15, Anderson19}), in addition to more diffuse emission associated with star formation. {At 241 Mpc, our $5\sigma$ sensitivity corresponds to a radio luminosity of $4\times10^{27}$ erg s$^{-1}$ Hz$^{-1}$ in epoch 1 and $6\times10^{27}$ erg s$^{-1}$ Hz$^{-1}$ in epoch 2. This is sufficient to detect all known radio TDEs \citep{Alexander20} and the brightest $\sim7\%$ of core-collapse supernovae \citep{Bietenholz21}.} Given the small number of galaxies we observed, their fairly proximal distances, and the rates of transients per $L_*$ galaxy {\citep{Li+11,metzger15,vv20}}, we expect to find $< 1$ serendipitous supernova or TDE in our search. However, our observations demonstrate that the contamination risk from AGN is much higher for observations with a cadence of several months, in agreement with previous work that shows radio AGN flares have a typical variability timescale of months to a few years (\citealt{hov08}; {\citealt{Richards+11}}).

Many galaxy-targeted searches for GW counterparts, including ours, use a prioritization scheme that ranks galaxies in part by their optical luminosity, with the intent of maximizing the fraction of stellar mass in the localization volume that can be observed with a finite amount of telescope time (e.g.~\citealt{Gehrels16,Arcavi17b,Ducoin20}). For radio interferometers in which slew times can be significant, observing strategies typically maximize a combination of probability covered and slew \citep{Rana19}, but do not necessarily address the issue of heightened contamination in galaxies. By default, such galaxy-ranking algorithms also have the effect of prioritizing exactly the galaxies that are most likely to have detectable radio emission from unrelated processes such as star formation or AGN activity, even at distances of a few hundred Mpc. Indeed, we find that 23 of our target galaxies (31\% of our sample) have detected radio emission coincident with the galaxy nucleus, and 17 of these nuclear radio sources are initially flagged as variable using our transient search criteria (i.e., we find that 74\% of these nuclear sources exhibited significant variability on timescales of $5-6$ months). We therefore conclude that identifying radio counterparts in or near the nuclei of the most massive galaxies will be challenging for low-cadence observations -- particularly as most existing all-sky radio surveys are fairly shallow, so there are unlikely to be sufficiently deep pre-merger radio observations of the target galaxies to compare against.

We also find that for radio variability searches conducted with the VLA or another reconfigurable radio interferometer, the array configuration needs to be carefully considered when planning GW follow up and interpreting the data. As mentioned above, many of the most highly-ranked galaxies in a GW localization volume may have detectable radio emission from other, unrelated sources; thus, lower-resolution observations taken in the VLA's C and D configurations (such as our epoch 2 observations) may suffer from issues of source confusion. In addition, while the VLA is equally sensitive to emission from point-like unresolved sources in all configurations, it is more sensitive to diffuse emission in its most compact configurations. Thus, sources identified as point-like in our high-resolution epoch 1 A-configuration data may in fact have more extended components that contribute to the fitted flux density measured in our lower-resolution epoch 2 C-configuration data (even when forcing a point source fit), giving the appearance of variability. We see some evidence for this in our data, complicating our efforts to determine how much of the variability is intrinsic for these sources. This reinforces the importance of using variability selection criteria tailored for galaxy-targeted searches, and taking extra care near the nuclei of target galaxies where such extended emission is most likely to be detected. Deep template images of each galaxy in the relevant configuration(s) would be necessary to attempt to deconvolve a potential near-nuclear GW counterpart from the background variability of its host.

\section{Conclusions}\label{sec:conc}

We carried out the first galaxy-targeted search for a radio counterpart to a gravitational wave merger event, GW190814. Although we detected several transient or variable sources, all are consistent with AGN variability or are artificial variables created by the different $uv$ coverage of our two epochs of data, and are thus unlikely to be genuine radio counterparts to GW190814. Via additional monitoring of one initially promising candidate, we demonstrate that multi-frequency radio observations can help distinguish background AGN flares from bona fide radio GW counterparts, as they may have different spectral indices and/or require unphysical parameter values to fit the shape of the radio light curve in the context of relativistic jet or kilonova afterglow models. For the 75 galaxies that we observed, comprising 32\% of the stellar luminosity in the final localization volume, we can rule out a relativistic jet at the best-fit LIGO/Virgo viewing angle of $\sim$46$^{\circ}$ with isotropic-equivalent energies $E_{K}$ = $2\times10^{51}$, $5\times10^{51}$, and $5\times10^{52}$ erg propagating in an ISM-like constant density medium of $n\gtrsim$ {0.4, 0.3, 0.02} cm$^{-3}$ for a tophat jet model, or ${n \gtrsim 0.1}$, {0.03}, ${9 \times 10^{-4}}$ cm$^{-3}$ for a Gaussian jet model (assuming $\theta_{jet}=15^{\circ}$, $p=2.2$, $\epsilon_e=0.1$, and $\epsilon_B=0.01$). {These limits will change if different values are assumed for $\theta_{jet}$, $p$, and the microphysical parameters; thus, care must be taken when cross-comparing the limits set by different studies in the literature.}

Our results have a number of implications for future radio searches for gravitational wave counterparts and other radio transients. In particular, we find that searches specifically targeting nearby galaxies (where a compact object merger is most likely to occur) may encounter additional complexities. We identify a significantly higher fraction of variable radio sources in our galaxy-targeted search, in comparison to previous wide-field radio transient searches (e.g.~\citealt{Carilli+03,frail12, Mooley16, Radcliffe19, bhakta20, sarb20}). This is likely due to a combination of several factors, some of which are relevant for any galaxy-targeted search (e.g.~contamination from unrelated AGN variability), and some of which are specific to the VLA (e.g.~the complications incurred by the regular reconfiguration of the VLA, which results in a non-uniform dataset for any followup campaign lasting $\gtrsim3$ months).

As Advanced LIGO, Virgo, Kagra, and future generations of GW detectors continue to improve their sensitivity, wide-field and galaxy-targeted radio counterpart searches will also need to improve their sophistication. In the future, sensitive, high-resolution, wide-field radio facilities like ASKAP \citep{Dobie2019}, MeerKAT, and the Square Kilometer Array will likely be able to overcome some of the challenges encountered in this work (particularly if higher-frequency receivers are added). Nevertheless, our pilot study demonstrates that there is a niche for galaxy-targeted radio searches for nearby events with the VLA. Our observations were sufficient to rule out most plausible SGRB-like jet models for the galaxies we observed, with a modest investment of only 23 hours of telescope time. In the future, we would suggest adding additional epochs to enable at least two observations of each target galaxy per array configuration, to minimize the resolution effects encountered in this study and to better discriminate between AGN variability and the smooth rise and decline expected for a GW radio counterpart. To prevent the total time investment from being prohibitive, we also suggest that this strategy only be applied to events that are closer and/or better localized than GW190814, such that the total number of galaxies in the localization volume is $\lesssim50$. Such events will remain rare (we expect at most one such BNS or NSBH merger in O4; \citealt{predictions20}), but are nevertheless likely to be among the best-studied mergers. The increasing availability of deep template maps of the radio sky will also improve our ability to interpret the results of future radio counterpart searches, just as we found VLASS observations of our target galaxies to be useful in this work. Radio searches will remain the only way to discover electromagnetic counterparts to compact object mergers that occur in the daytime sky, whose optical and X-ray emission cannot be studied, and will thus remain an important tool for multi-messenger studies, despite their challenges.

\acknowledgments
K.D.A. is supported by NASA through the NASA Hubble Fellowship grant \#HST-HF2-51403.001-A awarded by the Space Telescope Science Institute, which is operated by the Association of Universities for Research in Astronomy, Inc., for NASA, under contract NAS5-26555. Support for G.S. in this work was provided by the NSF
through Student Observing Support award SOSP19B-001 from
the NRAO. The Fong Group at Northwestern acknowledges support by the National Science Foundation under grant Nos. AST-1814782 and AST-1909358. P.~S.~C. is grateful for support provided by NASA through the NASA Hubble Fellowship grant \#HST-HF2-51452.001-A and \#HST-HF2-51404.001-A awarded by the Space Telescope Science Institute, which is operated by the Association of Universities for Research in Astronomy, Inc., for NASA, under contract NAS5-26555.
B.M. is supported by NASA through the NASA Hubble Fellowship grant \#HST-HF2-51412.001-A awarded by the Space Telescope Science Institute, which is operated by the Association of Universities for Research in Astronomy, Inc., for NASA, under contract NAS5-26555. B.D.M acknowledges support from NSF AAG (grant no.~GG016244).

The National Radio Astronomy Observatory is a facility of the National Science Foundation operated under cooperative agreement by Associated Universities, Inc. The Pan-STARRS1 Surveys (PS1) and the PS1 public science archive have been made possible through contributions by the Institute for Astronomy, the University of Hawaii, the Pan-STARRS Project Office, the Max-Planck Society and its participating institutes, the Max Planck Institute for Astronomy, Heidelberg and the Max Planck Institute for Extraterrestrial Physics, Garching, The Johns Hopkins University, Durham University, the University of Edinburgh, the Queen's University Belfast, the Harvard-Smithsonian Center for Astrophysics, the Las Cumbres Observatory Global Telescope Network Incorporated, the National Central University of Taiwan, the Space Telescope Science Institute, the National Aeronautics and Space Administration under Grant No. NNX08AR22G issued through the Planetary Science Division of the NASA Science Mission Directorate, the National Science Foundation Grant No. AST-1238877, the University of Maryland, Eotvos Lorand University (ELTE), the Los Alamos National Laboratory, and the Gordon and Betty Moore Foundation. This paper includes data gathered with the 6.5 m Magellan Telescopes located at Las Campanas Observatory, Chile. This work was performed in part at the Aspen Center for Physics, which is supported by National Science Foundation grant PHY-1607611.

\clearpage
\startlongtable
\begin{deluxetable*}{lllrrrrr}
\tabletypesize{\footnotesize}
\tablecaption{Galaxies targeted in our observations.\label{tab:obs}}
\tablehead{
\colhead{Galaxy Name}	&	\multicolumn{2}{c}{Date observed (UT)} &	\colhead{RA (J2000)}	& \colhead{DEC (J2000)} &	 \colhead{Distance}	&	 \colhead{$z$}	&	 \colhead{$M_B$} 	\\
 & \colhead{Epoch 1} & \colhead{Epoch 2} & \colhead{(deg)} & \colhead{(deg)} & \colhead{(Mpc)} & & }
\startdata 
ESO 474-026	&	2019 Sep 22.31362	&	2020 Feb 29.79077	&	11.781363	&	-24.370647	&	244.25	&	0.0263	&	-21.98	\\
IC 1587	&	2019 Sep 22.34542	&	2020 Feb 29.82260	&	12.180364	&	-23.561686	&	252.33	&	0.0442	&	-21.83	\\
PGC 198197	&	2019 Sep 21.34325	&	2020 Mar 14.78605	&	12.091079	&	-25.126814	&	297.78	&	0.0661	&	-21.39	\\
PGC 198196	&	2019 Sep 21.33787	&	2020 Mar 14.78064	&	11.870618	&	-25.440655	&	267.39	&	0.0594	&	-21.32	\\
PGC 2864	&	2019 Sep 22.35535	&	2020 Feb 29.83249	&	12.25617	&	-23.811317	&	236.13	&	0.0525	&	-21.14	\\
ESO 474-035{$^a$}	&	2019 Sep 18.35343	&	2020 Mar 13.88200	&	13.173257	&	-25.733852	&	271.35	&	0.0605	&	-20.92	\\
ESO 474-041	&	2019 Sep 21.42925	&	2020 Mar 14.87205	&	13.601437	&	-25.464052	&	203.64	&	0.0506	&	-21.37	\\
PGC 787700	&	2019 Sep 21.39944	&	2020 Mar 14.84222	&	12.874431	&	-24.642494	&	274.46	&	0.0612	&	-20.58	\\
PGC 3264	&	2019 Sep 18.38777	&	2020 Mar 13.91635	&	13.806392	&	-26.321253	&	264.40	&	0.0417	&	-21.87	\\
PGC 2998	&	2019 Sep 18.32365	&	2020 Mar 13.85223	&	12.828166	&	-26.16806	&	285.63	&	0.0635	&	-20.95	\\
PGC 133715	&	2019 Sep 18.27303	&	2020 Mar 13.80159	&	12.441799	&	-26.443037	&	244.28	&	0.0543	&	-21.18	\\
PGC 773232	&	2019 Sep 18.33814	&	2020 Mar 13.86670	&	12.793847	&	-25.954172	&	278.58	&	0.0623	&	-20.68	\\
PGC 3231	&	2019 Sep 18.38321	&	2020 Mar 13.91178	&	13.704722	&	-26.371256	&	238.93	&	0.0531	&	-21.30	\\
PGC 2694	&	2019 Sep 22.30904	&	2020 Feb 29.78620	&	11.543399	&	-24.650192	&	220.30	&	0.0495	&	-20.87	\\
PGC 133716	&	2019 Sep 18.28752	&	2020 Mar 13.81612	&	12.364808	&	-26.538301	&	224.63	&	0.0504	&	-21.35	\\
PGC 792107	&	2019 Sep 22.31909	&	2020 Feb 29.79624	&	11.771893	&	-24.238703	&	292.34	&	0.0650	&	-20.85	\\
PGC 133702	&	2019 Sep 21.35322	&	2020 Mar 14.79598	&	12.2428	&	-25.69345	&	296.48	&	0.0658	&	-20.63	\\
PGC 797191	&	2019 Sep 22.34998	&	2020 Feb 29.82717	&	12.178233	&	-23.773075	&	253.14	&	0.0568	&	-20.43	\\
PGC 133717	&	2019 Sep 21.35778	&	2020 Mar 14.80055	&	12.320091	&	-26.219179	&	331.27	&	0.0732	&	-21.70	\\
PGC 3235434	&	2019 Sep 21.43384	&	2020 Mar 14.87662	&	13.788131	&	-25.455704	&	232.85	&	0.0516	&	-20.59	\\
2MASX J00485495-2504100        	&	2019 Sep 21.34785	&	2020 Mar 14.79063	&	12.22897	&	-25.069471	&	242.49	&	0.0540	&	-20.00	\\
PGC 786964	&	2019 Sep 21.39407	&	2020 Mar 14.83686	&	12.473102	&	-24.707197	&	235.84	&	0.0524	&	-20.03	\\
PGC 3235517	&	2019 Sep 21.33328	&	2020 Mar 14.77604	&	12.103154	&	-25.595707	&	272.38	&	0.0607	&	-20.26	\\
PGC 2875	&	2019 Sep 22.35991	&	2020 Feb 29.83707	&	12.311129	&	-23.858547	&	200.29	&	0.0440	&	-20.65	\\
IC 1588	&	2019 Sep 22.37520	&	2020 Feb 29.85237	&	12.740481	&	-23.557995	&	242.52	&	0.0540	&	-20.64	\\
PGC 788830	&	2019 Sep 21.41931	&	2020 Mar 14.86207	&	13.490571	&	-24.543142	&	296.32	&	0.0658	&	-20.77	\\
PGC 798968	&	2019 Sep 22.36528	&	2020 Feb 29.84245	&	12.643789	&	-23.618547	&	229.77	&	0.0512	&	-20.47	\\
PGC 133703	&	2019 Sep 22.32906	&	2020 Feb 29.80625	&	11.472051	&	-23.772461	&	236.10	&	0.0526	&	-21.02	\\
PGC 783013	&	2019 Sep 21.42385	&	2020 Mar 14.86664	&	13.65696	&	-25.067099	&	220.76	&	0.0499	&	-20.42	\\
2MASX 00494172-2503029	&	2019 Sep 21.38953	&	2020 Mar 14.83232	&	12.42386	&	-25.050814	&	261.42	&	0.0581	&	-19.74	\\
PGC 769203	&	2019 Sep 18.31829	&	2020 Mar 13.84685	&	12.800914	&	-26.313063	&	263.87	&	0.0587	&	-20.35	\\
PGC 2780	&	2019 Sep 22.33550	&	2020 Feb 29.81264	&	11.867466	&	-23.02301	&	209.90	&	0.0457	&	-21.22	\\
PGC 3123	&	2019 Sep 18.36336	&	2020 Mar 13.89193	&	13.299205	&	-26.093979	&	205.87	&	0.0456	&	-20.55	\\
PGC 798818	&	2019 Sep 22.36984	&	2020 Feb 29.84702	&	12.726862	&	-23.631887	&	316.80	&	0.0701	&	-21.17	\\
PGC 198205	&	2019 Sep 22.38513	&	2020 Feb 29.86230	&	12.540436	&	-23.280048	&	256.82	&	0.0575	&	-20.88	\\
PGC 101138	&	2019 Sep 22.41494	&	2020 Feb 29.89210	&	13.804421	&	-24.044033	&	261.76	&	0.0458	&	-21.16	\\
2MASX 00530427-2610148	&	2019 Sep 18.36792	&	2020 Mar 13.89650	&	13.267828	&	-26.170792	&	245.22	&	0.0545	&	-20.08	\\
PGC 787067	&	2019 Sep 21.40400	&	2020 Mar 14.84679	&	13.064037	&	-24.69873	&	224.21	&	0.0505	&	-19.82	\\
PGC 3000	&	2019 Sep 18.30281	&	2020 Mar 13.83139	&	12.838246	&	-26.98945	&	272.00	&	0.0608	&	-21.31	\\
PGC 198217	&	2019 Sep 18.28296	&	2020 Mar 13.81155	&	12.346334	&	-26.507498	&	299.45	&	0.0665	&	-20.90	\\
PGC 786999	&	2019 Sep 21.40938	&	2020 Mar 14.85217	&	13.269273	&	-24.704401	&	234.47	&	0.0519	&	-19.76	\\
PGC 773004	&	2019 Sep 21.37313	&	2020 Mar 14.81589	&	12.551639	&	-25.975159	&	272.37	&	0.0608	&	-19.93	\\
PGC 2947	&	2019 Sep 18.29745	&	2020 Mar 13.82601	&	12.666141	&	-26.813395	&	317.44	&	0.0703	&	-21.64	\\
PGC 198221	&	2019 Sep 18.29289	&	2020 Mar 13.82147	&	12.385612	&	-26.538588	&	301.73	&	0.0670	&	-20.89	\\
PGC 198201	&	2019 Sep 22.37976	&	2020 Feb 29.85692	&	12.637869	&	-23.295488	&	239.17	&	0.0532	&	-20.73	\\
PGC 774472	&	2019 Sep 21.32782	&	2020 Mar 14.77060	&	11.9691	&	-25.8414	&	246.68	&	0.0550	&	-20.08	\\
PGC 142558	&	2019 Sep 18.27759	&	2020 Mar 13.80616	&	12.332034	&	-26.476397	&	254.34	&	0.0567	&	-20.41	\\
PGC 133700	&	2019 Sep 21.41394	&	2020 Mar 14.85671	&	13.477119	&	-24.077032	&	210.96	&	0.0471	&	-20.72	\\
ESO 474-036	&	2019 Sep 22.39506	&	2020 Feb 29.87222	&	13.192388	&	-22.975018	&	236.97	&	0.0480	&	-21.94	\\
PGC 766121	&	2019 Sep 18.37785	&	2020 Mar 13.90642	&	13.358978	&	-26.599817	&	278.84	&	0.0624	&	-20.48	\\
PGC 771842	&	2019 Sep 18.32821	&	2020 Mar 13.85680	&	12.881038	&	-26.077213	&	295.12	&	0.0656	&	-20.05	\\
PGC 198252	&	2019 Sep 21.32326	&	2020 Mar 14.76606	&	11.7816	&	-25.66073	&	268.67	&	0.0598	&	-20.12	\\
2MASX 00511861-2620430        	&	2019 Sep 18.31373	&	2020 Mar 13.84227	&	12.827568	&	-26.345284	&	247.73	&	0.0551	&	-19.92	\\
PGC 198242	&	2019 Sep 22.30363	&	2020 Feb 29.78082	&	11.264906	&	-25.019766	&	275.32	&	0.0614	&	-20.58	\\
PGC 773323	&	2019 Sep 21.36318	&	2020 Mar 14.80596	&	12.4675	&	-25.94625	&	306.20	&	0.0679	&	-20.11	\\
PGC 3235862	&	2019 Sep 18.35799	&	2020 Mar 13.88655	&	13.3536	&	-25.8268	&	260.30	&	0.0579	&	-19.51	\\
PGC 198247	&	2019 Sep 21.36777	&	2020 Mar 14.81056	&	12.525684	&	-25.957939	&	338.83	&	0.0748	&	-20.93	\\
PGC 2993	&	2019 Sep 18.30737	&	2020 Mar 13.83597	&	12.808345	&	-26.461119	&	202.90	&	0.0449	&	-20.49	\\
PGC 777373	&	2019 Sep 21.38416	&	2020 Mar 14.82694	&	12.7184	&	-25.57706	&	226.34	&	0.0507	&	-19.46	\\
PGC 200164	&	2019 Sep 22.29907	&	2020 Feb 29.77624	&	11.054956	&	-24.327574	&	288.19	&	0.0641	&	-20.75	\\
PGC 3083	&	2019 Sep 18.37329	&	2020 Mar 13.90188	&	13.149672	&	-26.750933	&	211.01	&	0.0469	&	-20.51	\\
PGC 773284	&	2019 Sep 21.31789	&	2020 Mar 14.76068	&	11.555073	&	-25.950188	&	268.84	&	0.0599	&	-20.92	\\
PGC 3235913	&	2019 Sep 18.34351	&	2020 Mar 13.87208	&	12.9027	&	-25.94219	&	261.52	&	0.0582	&	-19.36	\\
PGC 3196	&	2019 Sep 22.40954	&	2020 Feb 29.88672	&	13.566429	&	-23.535162	&	262.21	&	0.0583	&	-20.83	\\
PGC 3093	&	2019 Sep 18.34807	&	2020 Mar 13.87665	&	13.194542	&	-25.671635	&	169.14	&	0.0391	&	-20.70	\\
PGC 773198	&	2019 Sep 21.37957	&	2020 Mar 14.82234	&	12.6371	&	-25.95719	&	293.77	&	0.0653	&	-19.66	\\
PGC 100480	&	2019 Sep 18.39395	&	2020 Mar 13.92271	&	23.594837	&	-32.835316	&	286.65	&	0.0638	&	-22.18	\\
PGC 198243	&	2019 Sep 22.32367	&	2020 Feb 29.80084	&	11.3958	&	-24.24854	&	231.52	&	0.0516	&	-19.82	\\
PGC 198225	&	2019 Sep 22.34006	&	2020 Feb 29.81722	&	12.174922	&	-23.368631	&	330.03	&	0.0729	&	-21.14	\\
PGC 3198	&	2019 Sep 22.40498	&	2020 Feb 29.88215	&	13.570972	&	-23.552662	&	213.56	&	0.0480	&	-20.85	\\
PGC 2939	&	2019 Sep 22.38969	&	2020 Feb 29.86687	&	12.639195	&	-23.016602	&	236.27	&	0.0526	&	-20.66	\\
PGC 198202	&	2019 Sep 22.39962	&	2020 Feb 29.87677	&	13.522819	&	-23.194635	&	266.81	&	0.0593	&	-21.13	\\
PGC 133698	&	2019 Sep 22.41953	&	2020 Feb 29.89667	&	14.254662	&	-23.837297	&	220.63	&	0.0499	&	-21.21	\\
PGC 773149	&	2019 Sep 18.33358	&	2020 Mar 13.86215	&	12.8157	&	-25.9609	&	300.61	&	0.0667	&	-19.53	\\
IC 1581	&	2019 Sep 21.31334	&	2020 Mar 14.75611	&	11.442945	&	-25.920193	&	222.59	&	0.0500	&	-21.10	\\
\enddata
\tablenotetext{a}{Galaxy selected for additional multi-frequency follow up, based on the presence of a variable radio source within 100 kpc discovered in our observations.}
\end{deluxetable*}

\facilities{VLA} 

\software{{afterglowpy \citep{afterglowpy},} CASA \citep{casa}, emcee {\citep{emcee}}, NumPy {\citep{numpy}}, pwkit {\citep{pwkit}}}

\bibliographystyle{aasjournal}
\bibliography{ms}

\end{document}